\documentclass[aps,prl,preprint,superscriptaddress]{revtex4-1}
\usepackage{enumitem}
\setlist[itemize]{align=parleft,left=0pt..1em}

\usepackage{gensymb}
\usepackage{amssymb}
\usepackage{amsmath}
\usepackage{textcomp}
\usepackage{placeins}

\usepackage{color}
\usepackage{csquotes}
\usepackage{rotating} 
\usepackage{gensymb}
\usepackage{float}
\usepackage{esint}
\usepackage{url}
\usepackage{braket}
\usepackage{threeparttable} 
\usepackage{multirow}
\usepackage{booktabs}

\usepackage{mathtools}  
\usepackage{array}
\usepackage{hyperref}
\hypersetup{
    unicode=true,            
    colorlinks=true,         
       citecolor    = blue,
       filecolor = blue,       
       linkcolor = blue,
       urlcolor  = blue,
}
\usepackage{xcolor}

\makeatletter

\newcommand{\Rmnum}[1]{\expandafter\@slowromancap\romannumeral #1@}
\makeatother

\begin{document}

\title{\textbf{Comment on:} "The interaction of neutrons with $^7$Be at BBN temperatures: Lack of Standard Nuclear Solution to the “Primordial $^7$Li Problem” by M. Gai \textit{et al.}}

\author{M. Paul}
\affiliation{Racah Institute of Physics, Hebrew University, Jerusalem, Israel 91904}

\author{R. Dressler}
\affiliation{Paul Scherrer Institute, Villigen, Switzerland}

\author{U. K\"{o}ster}
\affiliation{Institut Laue-Langevin, Grenoble, France}

\author{D. Schumann}
\affiliation{Paul Scherrer Institute, Villigen, Switzerland}
\vspace*{5mm}


\begin{abstract}

The  article recently published by M. Gai \textit{et al.} claims to reach conclusions from a collaborative experiment dedicated to the study of the $^7$Be($n,\alpha$) reaction. These claims were published with no authorization from key collaborators, including a PI of the experiment.  The Authors of the present Comment reject the conclusions of M. Gai et \textit{et al.} and condemn the scientific and ethical misconduct involved in their publication. A  formal Comment, similarly expressing the Authors’ rejection of these conclusions was  submitted  to EPJ Web of Conferences who published the above article.

\vspace{10mm} 
\end{abstract}

%
\maketitle



The publication \cite{gai20}  (see also \cite{gai18})  addresses the important reaction $^7$Be($n,\alpha$)  as possibly bearing consequences in the resolution of the so-called “Primordial Lithium Problem” \cite{FIEXX}. An experiment aimed at the study of the above reaction was initiated, lead by PI’s M.P. (Hebrew University of Jerusalem, HUJ) and M. Gai (University of Connecticut, UC) under the auspices of US-Israel Binational Science Foundation \cite{BSF14}.  The experiment involved a wide collaboration of researchers from UC, Soreq Nuclear Research Center (SNRC, Israel), Paul Scherrer Institute (PSI, Switzerland),  Institut Laue-Langevin (ILL, France),  HUJ,  Triangle Universities Nuclear Laboratories (TUNL, Duke University, USA), Weizmann Institute of Science (Israel), and CERN (Switzerland). The experiment itself was performed at the Soreq Applied Research Accelerator Facility (SARAF, SNRC) \cite{SARAF} using a $^7$Be target prepared specifically for this purpose at PSI and the Liquid-Lithium Target \cite{RSI,EPJA} designed and built at SARAF by a HUJ-SARAF  collaboration. The 
SARAF-LiLiT setup is known to produce an intense yield ($\sim2\times10^{10}$ n/s) of quasi-Maxwellian neutrons in the range $kT=30-40$ keV depending on the geometrical conditions, well suited for the study of neutron induced reactions relevant for some of the astrophysical sites. Decision was made to use CR-39 solid-state track detectors for the detection of $\alpha$ particles and discrimination from abundant protons from the $^7$Li($n,p$) reactions in the presence of  the intense neutron and gamma background and electromagnetic noise from the LiLiT apparatus. A preparation experiment designed to observe $\alpha$ particle tracks produced by neutrons from  the $^{10}$B($n,\alpha$)$^7$Li  was performed at SARAF-LiLiT by the HUJ-SARAF groups in the frame of the BSF project. The experiment successfully demonstrated the feasibility of detection of $\alpha$ particles in this hostile environment with semi-quantitative results, nonetheless justifying a first experiment with a $^7$Be target.  A number of calibration experiments were designed and performed by the UC group to characterize $\alpha$ and proton pits in a CR-39 detector. Irradiation runs with quasi-Maxwellian neutrons produced at LiLiT by the intense  ($\sim$1 mA) proton beam at 1.94 MeV from SARAF were designed and performed on targets of  $^{10}$B (for proof of principle), $^7$B and $^9$Be (background) targets under different experimental conditions. Etching and scanning of the CR-39 plates were performed in part at HUJ and UC and for the scanning also at Bar-Ilan University (in collaboration with A. Weiss, HUJ). 

In spite of the methodology described above, the Authors of the present Comment (the Authors) are of the opinion that the SARAF-LiLiT failed to reach reliable or quantitative results. This assessment was repeatedly  expressed to M. Gai by the Authors and spelled out by Schumann et al. \cite{sch19}.  We list below the severe flaws leading to our rejection of the conclusions of [1].

1. The calibration experiments for $\alpha$ particles detected in the CR-39 track detectors presented by M. Gai and collaborators in \cite{gai20} and in other Conference contributions \cite{kad15,kad16,gai17} display notable discrepancies. The source of these discrepancies is not clear and the arbitrary choice of one set among the others, not confirmed by any additional independent calibration experiment, is  considered unacceptable by the Authors and unfit for extraction of physical information. 

2. The background present in the spectra of pit radii attributed to alpha particles from the $^7$Be($n,\alpha$)  reaction in the SARAF-LiLiT experiment (Fig. 2 in \cite{gai20}) is considered by the Authors unfit to identify a reliable signal and to extract quantitative information on the reaction cross section. 

3. M. Gai \textit{et al.} claim in \cite{gai20} to identify protons emitted in the same experiment by the $^7$Be($n,p$)$^7$Li reaction and to determine the cross section of this reaction, based on a calibration shown in their Fig. 3. The spectrum of pit radii taken as calibration for protons in their Fig. 3 is however a steep decreasing function of the radius. In these conditions, the interval [0.8 $\mu$m, 1.4 $\mu$m]  taken by Gai \textit{et al.} as the radii region of interest for protons cannot lead to a quantitative determination of proton-induced pits. In fact, a recent publication of the same group \cite{KAD20} establishes the uncertainty of pit radius determination at 0.2 $\mu$m, caused by temperature variations during the etching procedure: this uncertainty leads to a change of one order of magnitude in assigned pits owing to the steepness of the calibration curve.  A low efficiency of proton counting of 8.7\%, determined with ambiguous uncertainty  of 3\% (relative uncertainty of 34\%) \cite{gai20} and 1.3\% (relative uncertainty of 14\%) \cite{KAD20}, compounds the problem. The Authors reject the quantitative extraction of the $^7$Be($n,p$) cross section claimed by Gai \textit{et al.} in \cite{gai20}.

The Authors express again and unequivocally their position that the above experiment failed to achieve its goals and that in no circumstance it can be used to extract quantitative information on a physical quantity such as a reaction cross section. The flawed approach taken by M. Gai in the handling, interpretation and presentation of data culminated in his unethical publication of results rejected and disavowed by key Collaborators. A normal course of action following an unsuccessful experiment or conflicting opinions between co-experimenters, neither an infrequent circumstance, is identifying and correcting their likely causes and having a repeat trial. We regret that irreconcilable standpoints and unilateral steps of M. Gai rendered collaboration with him unfeasible. They led a PI (M.P.) to withdraw from collaboration in his own experiment and key members to distance themselves from further publications and prevented the pursuit of an interesting scientific project.

\bibliography{Comment_v2
}


\end{document}